\documentstyle[sprocl,epsfig]{article}

\def\be{\begin{equation}}
\def\ee{\end{equation}}
\def\bea{\begin{eqnarray}}
\def\eea{\end{eqnarray}}

\newcommand{\case}[2]{\mbox{\footnotesize $\displaystyle \frac{#1}{#2}$}}
\newcommand{\lsim}{\mathrel{\rlap{\lower3pt\hbox{\hskip0pt$\sim$}}
\raise2pt\hbox{$<$}}}
\newcommand{\gsim}{\mathrel{\rlap{\lower3pt\hbox{\hskip0pt$\sim$}}
\raise2pt\hbox{$>$}}}

\begin{document}

\title{CONTEMPORARY APPLICATIONS OF\\  DYSON-SCHWINGER EQUATIONS\\[-1ex]}

\author{M.\ B.\ HECHT, C.\ D.\ ROBERTS and S.\ M.\ SCHMIDT}

\address{Physics Division, Bldg. 203, Argonne National Laboratory\\
Argonne IL 60439-4843, USA}


\maketitle
\abstracts{
 \begin{center}
 \parbox{30em}{\sc Contribution to the Proceedings of ``Confinement IV: The
 4th International Conference on Quark Confinement and the Hadron
 Spectrum,'' 3-8/Jul./2000, Vienna, Austria.\\[0.5ex]} \end{center}  
We illustrate the contemporary application of Dyson-Schwinger equations using
two examples: the calculation of pseudoscalar meson masses, an associated
model-independent mass formula and the approach to the heavy-quark limit; and
the study of nucleon observables, including a calculation of its mass, $M$,
via a covariant Fadde'ev equation and an estimate of pion-loop contributions
to $M$.
}\vspace*{-2ex}

\hspace*{-\parindent}{\bf 1.~~~Introductory Remarks.}~~The Dyson-Schwinger
equations (DSEs) in quantum field theory are an analogue of the
Euler-Lagrange equations in classical field theory.  They are an enumerable
infinity of coupled integral equations whose solutions are the $n$-point
Schwinger functions (Euclidean Green functions).  These Schwinger functions
are also the matrix elements estimated in numerical simulations of
lattice-QCD.  The coupling between equations poses a challenge, of course: it
necessitates a truncation in order to define a tractable problem.  A weak
coupling expansion provides one systematic method and reproduces perturbation
theory.  However, it also makes nonperturbative phenomena inaccessible, and
something else is needed for the study of strongly interacting systems and
bound state phenomena.

This situation is familiar from many body theory where the Hartree-Fock
truncation often yields reliable information.  There is an analogue in QCD:
the renormalisation-group-improved rainbow-ladder truncation, which is
phenomenologically efficacious, as is clear from Ref.\ [\ref{marisviennaR}]
wherein the power of a single-parameter model of the infrared behaviour of
the effective quark-quark interaction is illustrated.\footnote{The
ultraviolet behaviour is fixed and model-independent because this truncation
of the DSEs reproduces perturbation theory.}  The successes and failures of
such a model can be understood once it is appreciated that the rainbow-ladder
truncation is the leading order in a systematic, Ward-Takahashi identity
preserving $1/N_c$-like expansion.\cite{truncscheme} But this demonstration
is a single, small step toward a rigorous foundation for contemporary DSE
modelling.

One of the beauties of a model is that its simplicity makes possible a rapid
comparison between theory and experiment.  Another is that it can be wrong:
an {\it Ansatz} is intuitively motivated and explored, and, if it is widely
successful, its failures can point to unanticipated phenomena.  Irrespective
therefore of the difficulties that remain in providing a rigorous foundation
for the application of DSEs in QCD, their phenomenological application plays
an important part in attempts to elucidate nonperturbative phenomena.

This is nowhere clearer than in the study of dynamical chiral symmetry
breaking (DCSB).  The DSEs provide the simplest and most direct means of
understanding the dichotomy of the pion as both a Goldstone mode and a bound
state of a massive dressed-quark and -antiquark,\cite{mrt98,mr97} and the
result is model-independent.  The analysis relies on the interplay between
the QCD gap equation, the DSE for the dressed-quark propagator, and the
inhomogeneous Bethe-Salpeter equation (BSE) for the axial-vector vertex.
This interplay is a consequence only of the axial-vector Ward-Takahashi
identity.  Furthermore, a key quantitative conclusion
follows:$\,$\cite{fredIRnew,cdrwienI} in order to reproduce observed
characteristics of the spectrum, the kernel in the QCD gap equation must
exhibit a significant enhancement on the domain $\Lambda_{\rm QCD}^2\lsim
k^2\lsim 2\,$GeV$^2$. Identifying the origin of that enhancement; i.e.,
whether it is a feature of the dressed-gluon propagator alone or of the
contraction of this propagator with the dressed-quark-gluon vertex, is
currently an important focus.\cite{cdrwienI}$^{-}$\cite{williamsvienna}

In the last decade the use of DSEs in QCD has attracted renewed interest and
they have been applied to a broad range of phenomena, as is clear from the
detailed summaries in Refs.\ [\ref{revbastiR},\ref{revreinhardR}].  The
approach has moved far beyond the calculation of indigent approximations to
$f_\pi$ and the vacuum quark condensate.  Herein we illustrate this by
focusing on two topics: pseudoscalar meson masses and their evolution with
the current-quark mass; and a description of the nucleon, its interactions
and the calculation of its mass.

\medskip

\hspace*{-\parindent}{\bf 2.~~~Pseudoscalar Meson Masses.}~~Meson masses can
be calculated by solving the renormalised homogeneous BSE:
\begin{eqnarray}
\label{genbse}
\left[\Gamma_H(k;P)\right]_{tu} &= & 
\int^\Lambda_q  \,
[\chi_H(q;P)]_{sr} \,K^{rs}_{tu}(q,k;P)\,,
\end{eqnarray}
where: $r$,\ldots,$u$ represent colour-, Dirac- and flavour-matrix indices
and $H$ identifies the meson under consideration; $P$ is the total momentum
and $P^2=-m_H^2$ is the eigenvalue condition for a solution; $\chi_H(q;P) :=
{\cal S}(q_+)\, \Gamma_H(q;P)\,{\cal S}(q_-)$ is the Bethe-Salpeter wave
function, with $\Gamma_H(q;P)$ the fully-amputated Bethe-Salpeter amplitude
and ${\cal S}={\rm diag}(S_u,S_d,S_s,\ldots)$ the dressed-quark propagator
flavour matrix, $q_+=q + \eta_P\, P$, $q_-=q - (1-\eta_P)\,
P$;$\,$\footnote{$\eta_P\in [0,1]$ is the momentum partitioning parameter.
It appears because in a Poincar\'e covariant approach the definition of the
relative momentum is arbitrary.  Observable quantities must be independent of
$\eta_P$.}
and $K(q,k;P)$ is the renormalised, fully-amputated quark-antiquark
scattering kernel, which is two-particle-irreducible, with respect to the
quark-antiquark pair of lines, and does not contain quark-antiquark to single
gauge-boson annihilation diagrams, such as would describe the leptonic decay
of a pseudoscalar meson.  In Eq.\ (\ref{genbse}) $\int^\Lambda_q :=
\int^\Lambda d^4 q/(2\pi)^4$ represents mnemonically a {\em
translationally-invariant} regularisation of the integral, with $\Lambda$ the
regularisation mass-scale, and in this particular case the r.h.s. of the
renormalised equation is cutoff-independent.  ($\Lambda\to \infty$ is the
final step in all calculations.)  Equation (\ref{genbse}) is obtained from
the inhomogeneous BSE by equating meson pole residues.

For a pseudoscalar meson the solution of Eq.\ (\ref{genbse}) has the form
\begin{eqnarray}
\lefteqn{\displaystyle\Gamma_H(q;P)  =}\\
&& \nonumber 
T^H \gamma_5 \left[\rule{0mm}{2.4ex} i E_H(q;P) + \gamma\cdot P F_H(q;P) +
\gamma\cdot q \, G_H(q;P) + \sigma_{\mu\nu}\,q_\mu P_\nu \,H_H(q;P)
\right]\,,
\end{eqnarray}
where $T^H$ is the flavour matrix that specifies the mesonic channel under
consideration; e.g., $T^{K^+} = \mbox{\small $(1/\surd 2)$}\left(\lambda^4 +
i \lambda^5\right)$, with $\{\lambda^i,i=1,\ldots,8\}$ the Gell-Mann
matrices.  The amplitude is canonically normalised by requiring that the
bound state contribution to the fully-amputated quark-antiquark scattering
matrix: $M = K + K ({\cal S}{\cal S}) K + \ldots\,$, have unit residue.
(See; e.g., Ref.\ [\ref{mr97R}].)

Using: the inhomogeneous BSEs for the axial-vector and pseudovector vertices;
the dressed-quark DSE; and the fact that a nonperturbative Ward-Takahashi
identity preserving truncation of the DSEs is possible, it was shown in Ref.\
[\ref{mrt98R}] that, for flavour nonsinglet pseudoscalar mesons,
\begin{equation}
\label{gmor}
\displaystyle f_H m_H^2 = {\cal M}_H^\zeta r_H^\zeta\,,
\end{equation}
with ${\cal \cal M}_H^\zeta:= {\rm tr}_{\rm flavour}
[M_{(\zeta)}\,\{T^H,(T^H)^{\rm t}\}]$\,, where $M_{(\zeta)}={\rm
diag}(m_u^\zeta,m_d^\zeta,m_s^\zeta,\ldots)$ and $(\cdot)^{\rm t}$ indicates
matrix transpose, so that ${\cal \cal M}_H^\zeta$ is proportional to the sum
of the constituents' current-quark masses.  This model-independent identity
is valid for all current-quark masses, irrespective of their magnitude, and
therefore provides a single formula that unifies the light- and heavy-quark
regimes.

In Eq.\ (\ref{gmor}), $f_H$ is the leptonic decay constant, which the
derivation proves is given by
\begin{equation}
\label{fH}
f_H\, P_\mu = Z_2\int^\Lambda_q\,\case{1}{2} {\rm
tr}\left[\left(T^H\right)^{\rm t} \gamma_5 \gamma_\mu \chi_H(q;P)\right]\,,
\end{equation}
where $Z_2=Z_2(\zeta,\Lambda)$ is the dressed-quark wave function
renormalisation constant, with $\zeta$ the renormalisation point.  This
multiplicative factor of $Z_2$ on the r.h.s.\ ensures that $f_H$ is
\underline{gauge-invariant}, and independent of $\zeta$ and $\Lambda$; i.e.,
that it is an observable.\cite{mrt98} Equation (\ref{fH}) yields $f_H$ as the
pseudovector projection of the meson's Bethe-Salpeter wave function at the
origin in configuration space; i.e., this equation provides a field
theoretical analogue of the ``wave function at the origin,'' which describes
the decay of bound states in quantum mechanics.

The remaining term in Eq.\ (\ref{gmor}) is
\begin{equation}
\label{rH}
i r_H^\zeta\, = Z_4\int^\Lambda_q\,\case{1}{2} {\rm
tr}\left[\left(T^H\right)^{\rm t} \gamma_5 \chi_H(q;P)\right]\,,
\end{equation}
where $Z_4=Z_4(\zeta,\Lambda)$ is the dressed-quark mass renormalisation
constant.  The gauge dependence of $Z_4$ is precisely that necessary to
ensure that the r.h.s.\ of Eq.\ (\ref{gmor}) is \underline{gauge invariant};
its cutoff dependence ensures that the r.h.s.\ is independent of the cutoff;
and its renormalisation point dependence ensures that the product on the
r.h.s.\ is independent of the renormalisation point.\cite{mrt98} $r_H^\zeta$
is the pseudoscalar projection of the meson's Bethe-Salpeter wave function at
the origin in configuration space.

In asymptotically free theories the chiral limit is unambiguously
de\-fined$\,$\cite{mrt98,mr97} by setting $\hat m=0$, where $\hat m$ is the
renormalisation point independent current-quark mass.  In this
limit$\,$\cite{mrt98}
\begin{equation}
\label{rH0}
r_{H_0}^\zeta := \lim_{\hat m \to 0}\,r_H^\zeta = 
-\frac{1}{f_{H_0}}\,\langle \bar q q \rangle_\zeta^0\,,
\end{equation}
where $f_{H_0}$ is obtained by taking the chiral limit in Eq.\ (\ref{fH}) and
\begin{equation}
\label{qbq0}
-\langle \bar q q \rangle_\zeta^0 = 
Z_4\,N_c\,\int_q^\Lambda {\rm tr}_D\left[S_{\hat m=0}(q)\right]  \,,     
\end{equation}
with $S_{\hat m=0}$ obtained as the chiral limit solution of the
dressed-quark DSE:
%
%
this is the gauge-invariant and cutoff-independent expression for the vacuum
quark condensate.  Using Eq.\ (\ref{rH0}), Eq.\ (\ref{gmor}) yields
\begin{equation}
\label{gmorO}
f_{H_0}^2 m_H^2 = - ( m_{f_1}^\zeta + m_{f_2}^\zeta )\,
\langle \bar q q \rangle_\zeta^0 + {\rm O}(\hat m_{f_1,f_2}^2)\,,
\end{equation}
with $f_{1,2}$ labelling the flavour of the dressed-quark constituents; i.e.,
as a corollary, Eq.\ (\ref{gmor}) yields the so called
Gell-Mann--Oakes--Renner relation.

As remarked above, Eq.\ (\ref{gmor}) is also valid for arbitrarily large
current-quark masses and an analysis of its heavy-quark limit is facilitated
by writing $P_\mu = m_H \,v_\mu = (\hat M_Q + E_H)$, where $\hat M_Q$ is a
constituent-heavy-quark mass$\,$\cite{mishaplb} and $E_H$ is a ``binding
energy.''  Following this the dressed-propagator for the heavy-quark
constituent takes the form
\begin{equation}
\label{hqf}
S_Q(k+P) = \case{1}{2}\,\frac{1 - i \gamma\cdot v}{k\cdot v - E_H}
+ {\rm O}\left(\frac{|k|}{\hat M_{Q}},
                \frac{E_H}{\hat M_{Q}}\right)\,,
\end{equation}
where $k$ is the momentum of the lighter constituent, and the canonically
normalised Bethe-Salpeter amplitude can be expressed as
\begin{equation}
\label{hqG}
\Gamma_H(k;P) = \sqrt{m_H}\,\Gamma_H^{<\infty}(k;P)\,,
\end{equation}
where $\Gamma_H^{<\infty}(k;P)$ is pointwise finite in the limit $m_H\to
\infty$.  Using Eqs.\ (\ref{hqf}) and (\ref{hqG}) in Eq.\ (\ref{fH})
yields$\,$\cite{mishaplb}
\begin{equation}
\label{fHHQ}
f_H = \frac{c_H^f}{\sqrt{m_H}}\,,
\end{equation}
with $c_H^f$ a calculable and finite constant, which reproduces a well-known
consequence of heavy-quark symmetry.  Applying the same analysis to Eq.\
(\ref{rH}) one finds$\,$\cite{marisAdelaide}
\begin{equation}
r_H^\zeta = c_H^{r_\zeta}\,\sqrt{m_H}
\end{equation}
and this, along with Eq.\ (\ref{fHHQ}) in Eq.\ (\ref{gmor}),
proves$\,$\cite{marisAdelaide,mishaSVY} that in the heavy-quark limit
\begin{equation}
\label{mHHQ}
m_H = \mbox{\small $\displaystyle\frac{c_H^{r_\zeta}}{c_H^f}$}\,{\cal
M}_H^\zeta\,;
\end{equation}
i.e., that the mass of a heavy pseudoscalar meson rises linearly with the
mass of its heaviest constituent.

It has been shown$\,$\cite{mishaSVY} that Eq.\ (\ref{fHHQ}) is not valid
until current-quark masses $m \gsim m_b$.  The $c$-quark lies well-outside
this domain; e.g., if a constant of proportionality is chosen so as to
reproduce the value of $f_B$, then $f_D$ obtained from this formula is $\sim$
40\% too large.  This is consistent with the calculated magnitude of the
violations of heavy-quark symmetry in $b\to c$ transitions ($\lsim
30$\%).

Using the results reported in Ref.\ [\ref{marisviennaR}], obtained using the
renormalisation-group-improved rainbow-ladder truncation of Ref.\
[\ref{pieterVMR}], one can study the evolution of meson masses as the
current-quark mass is increased.  Consider first pseudoscalar mesons whose
constituents have equal current-masses, for which the calculated evolution is
described by the interpolating formula$\,$\cite{marisvienna}
\begin{equation}
\label{mHX}
m_{H^\pi} = \beta\, \sqrt{{\cal X}} + \gamma\, {\cal X}\,,
\end{equation}
where $\beta=1.04\,$GeV, $\gamma=0.21\,$GeV and ${\cal X}=
m^{\zeta}/\Lambda_{\rm QCD}$, with $\zeta=19\,{\rm GeV}$ and $\Lambda_{\rm
QCD}=0.234$.  This formula was determined via an unconstrained fit to the
masses calculated by solving the Bethe-Salpeter equation.  A comparison with
Eq.\ (\ref{gmorO}) shows that $B_0^\zeta:=\beta^2/(2 \Lambda_{\rm QCD}) = -
\langle \bar q q \rangle_\zeta^0/f_{H_0}^2$ and, using the model's calculated
value of $f_{H_0}=0.088\,$GeV, one infers a value of $\langle \bar q q
\rangle_{\zeta}^0 = (-0.26\,{\rm GeV})^3$ from this correspondence.  That can
be compared with the value $(-0.27)^3$ calculated directly in this model from
Eq.\ (\ref{qbq0}).  This near equality indicates that the interpolating
formula in Eq.\ (\ref{mHX}) can provide reliable estimates.

Equation (\ref{mHX}) indicates that a flavour nonsinglet, $f_1=f=f_2$
pseudoscalar meson with a mass $m_H=1\,$GeV would be composed of quarks with
mass $\hat m_f = 0.32\,\Lambda_{\rm QCD} = 2.3\,\hat m_s$.  At this
current-quark mass, which corresponds to ${\cal X}= 0.68$, the $\sqrt{\cal
X}$ term still provides $86$\% of the meson's mass.  Thus one remains well
away from the linear trajectory, in spite of the fact that in the
neighbourhood of ${\cal X}= 0.68$ a tangent to the curve in the left panel of
Fig.~2, Ref.\ [\ref{marisviennaR}], is nearly indistinguishable from the
curve itself within the resolution of that figure.  Furthermore one finds
easily from Eq.\ (\ref{mHX}) that
\begin{equation}
\frac{m^2_{H^\pi_{m_f = 2 m_s}}}{m^2_{H^\pi_{m_f = m_s}}} = 2.2
\end{equation}
in agreement with the result obtained in recent lattice
simulations.\cite{cmichael} Hence, in addition to being phenomenologically
efficacious, the renormalisation-group-improved rainbow-ladder truncation of
Ref.\ [\ref{pieterVMR}] predicts a mass-evolution that is confirmed by
lattice simulations.  These results support a scenario of DCSB in which the
vacuum quark condensate is large; i.e., $B_0^{1\,{\rm GeV}}\gg f_{H_0}$.


Reference [\ref{marisviennaR}] also provides an interpolation of the
kaon-like $u$-$q$ trajectory: 
\begin{equation}
\label{mHK}
m_{H^K} = 0.083 + 0.5 \,\sqrt{\cal X} + 0.31\, {\cal X}, 
\end{equation}
and while the BSE studies reviewed in Ref.\ [\ref{pieterVMR}] have not
directly addressed heavy-light nor heavy-heavy bound states (systems with
$\hat m_f > 3.5 \,\hat m_s$ have not been studied) one may, as a preliminary
step, ask whether Eq.\ (\ref{mHK}) can be used to obtain reliable
mass-estimates via extrapolation?  Using Eq.\ (\ref{mHK}), one reproduces
$m_D\simeq 1.9\,$GeV, $m_B\simeq 5.3\,$GeV with $m_c^{1 \rm GeV} \simeq
1.1\,$GeV, $m_b^{1 \rm GeV} \simeq 4.2\,$GeV, and since these current-quark
masses are in agreement with other estimates$\,$\cite{pdg00} then Eq.\
(\ref{mHK}) can be a useful tool.  In this application one finds that the
linear term provides $50$\% of $m_D$ and $67$\% of $m_B$.  Thus, like the
conclusion drawn on the validity of Eq.\ (\ref{fHHQ}) for $f_H$, the
heavy-quark limit in Eq.\ (\ref{mHHQ}) is not valid until the current-quark
masses satisfy $m \gsim m_b$, and the $c$-quark mass is much smaller than
this lower bound.

\medskip

\hspace*{-\parindent}{\bf 3.~~~A Model of the Nucleon.}~~The success of the
rainbow-ladder truncation in describing meson observables
motivates$\,$\cite{regfe} a treatment of the nucleon as a bound state of a
dressed-quark and nonpointlike diquark via a covariant Fadde'ev equation.
The feasibility of this approach was demonstrated in Ref.\ [\ref{conradfeR}]
and the most extensive study to date is described in Ref.\ [\ref{reinhardR}].
The approach assumes only that the colour-$\bar 3$ quark-quark scattering
matrix can be approximated by a sum of diquark pseudoparticle terms: scalar
$+$ pseudovector $+$ \ldots, whose properties can be determined
independently.  The Fadde'ev equation then describes the nucleon as a
quark-diquark composite, which is bound by the repeated exchange of roles
between the dormant and diquark-participant quarks, and the complete nucleon
amplitude is a sum of three terms:
\begin{equation}
\label{Psi}
\Psi= \Psi_1 + \Psi_2+\Psi_3\,,
\end{equation}
where the subscript identifies the dormant quark and; e.g., $\Psi_{1,2}$ are
obtained from $\Psi_3$ via a cyclic permutation of the indices: $\Psi_1 =
\Psi_3(\downarrow^{123}_{231})$, $\Psi_2=\Psi_3(\downarrow^{123}_{312})$.

The simplest such model retains only the contribution of the scalar diquark
to the quark-quark scattering matrix, in which case
\begin{eqnarray}
\nonumber
\lefteqn{\Psi_3(p_i;\alpha_i,\tau_i;\alpha,\tau)=}\\
& & 
\epsilon_{c_1 c_2 c_3}\,\delta^{\tau\tau_3}\,\Delta^{0^+}(K)\,
[\Gamma_{0^+}(\case{1}{2}p_{[12]};K)]_{\alpha_1\alpha_2}^{\tau_1 \tau_2}\,
\psi_3(\ell;P)\,u(P)\,,
\label{Psi3}
\end{eqnarray}
where: $(i\gamma\cdot P + M) u(p)=0$, with $P=p_1+p_2+p_3=:p_{\{123\}}$ the
nucleon's total momentum and $M$ its mass; $\epsilon_{c_1 c_2 c_3}$ is the
colour-singlet factor; $K=p_1+p_2=:p_{\{12\}}$, $p_{[12]}:=p_1-p_2$, $\ell=
(p_{\{12\}} - 2 p_3)/3$; $(\alpha_i,\tau_i)$ are the quark spinor and isospin
labels, and $(\alpha,\tau)$ are those of the nucleon.  In Eq.\ (\ref{Psi3}),
$\Delta^{0^+}$ is the pseudoparticle propagator for a scalar diquark formed
from quarks $1$ and $2$, and $\Gamma_{0^+}$ is a Bethe-Salpeter-like
amplitude describing their relative momentum correlation.  Both these
quantities are determined by studying the quark-quark scattering matrix.  The
remaining element, $\psi_3$, is a Bethe-Salpeter-like amplitude that
describes the relative momentum correlation between the third quark and the
diquark's centre-of-momentum.  It satisfies a renormalised Fadde'ev equation,
which in the isospin symmetric limit assumes the form
\begin{eqnarray}
\nonumber \psi_3(k;P)\,u(P) & = & -2
\int_\ell^\Lambda\,\Delta^{0^+}(K_\ell)\,\Gamma_{0^+}(k+\ell/2;K) \\ & &
\times \, S(\ell_{\rm ex})^{\rm
t}\,\bar\Gamma_{0^+}(\ell+k/2;-K_k)\,S(\ell_1) \,\psi_3(\ell;P)\,u(P) \,,
\label{faddeev}
\end{eqnarray}
with $K_\ell= -\ell + (2/3)P$, $\ell_{\rm ex}= -\ell-k-P/3$, $\ell_1=\ell +
P/3$.  The general solution for a positive energy nucleon takes the form
\begin{equation}
\label{psi3}
\psi_3(\ell;P)= f_1(\ell;P) 1\!\rule{0.3ex}{1.55ex} - \frac{1}{M} \left( i
\gamma\cdot \ell - \ell\cdot \hat P \, 1\!\rule{0.3ex}{1.55ex}\right)
f_2(\ell;P)\,, 
\end{equation}
with $\hat P^2 = -1$ and where, in the nucleon rest frame, $f_{1,2}$
describe, respectively, the upper$/$lower component of the dressed-nucleon
spinor.

The nucleon amplitude in Eqs.\ (\ref{Psi}), (\ref{Psi3}) has been used
successfully to calculate a wide range of leptonic and nonleptonic nucleon
form factors.$\,$\cite{jacques1,jacques2} In those calculations, solving the
Fadde'ev equation was side-stepped by employing simple parametrisations for
the functions in Eqs.\ (\ref{Psi3}), (\ref{psi3}):
\begin{eqnarray}
\Delta^{0^+}(K^2) & = & \frac{1}{m_{0^+}^2}\,{\cal
F}(K^2/\omega_{0^+}^2)\,,\\
\label{gdq}
\Gamma_{0^+}(k;K) & = & \frac{1}{{\cal N}_{0^+}} C i\gamma_5\, i\tau_2\,
{\cal F}(k^2/\omega_{0^+}^2)\,,\\
f_1(\ell;P) & = & \frac{1}{{\cal N}_\psi} {\cal F}(\ell^2/\omega_{\psi}^2)\,,
\end{eqnarray}
with $f_2 = \mbox{\sc r} f_1$, where {\sc r} is a constant of proportionality
that gauges the relative importance of the lower component of the nucleon
spinor.  In these equations $C=\gamma_2\gamma_4$ is the charge conjugation
matrix, ${\cal F}(y) = (1- {\rm e}^{-y})/y$, and ${\cal N}_\psi$, ${\cal
N}_{0^+}$ are calculated, canonical normalisation constants.

The parameters in this model are $\omega_{0^+}$, $m_{0^+}$, $\omega_\psi$:
$d_{0^+}=1/\omega_{0^+}$ measures the quark-quark separation in the diquark;
$\ell_{0^+}=1/m_{0^+}$ is the diquark correlation length or mean free path;
and $d_\psi=1/\omega_\psi$ measures the quark-diquark
separation.$\,$\footnote{A description of the nucleon in this form can only
be internally consistent if $d_{0^+} < d_\psi$ and $\ell_{0^+}< d_\psi$;
i.e., the diquark is smaller than the nucleon and can't propagate over
distances larger than the nucleon.}  Their values have been determined by
requiring a good impulse-approximation fit to the proton's charge form factor
on $Q^2\in[0,3]\,$GeV$^2$ and this procedure yields
\begin{equation}
\label{paramsNFF}
\begin{array}{l|ccc}
\mbox{in GeV}   & \;\omega_\psi \;& \;\omega_{0^+} \;&\; m_{0^+} \\\hline
\mbox{\sc r}=1  &  \;0.19\;  &\; 0.68 \;  &\; 0.64 \\
\mbox{\sc r}=0.65  &  \;0.21\;  &\; 0.72 \;  &\; 0.63
\end{array}\,,\;\;\;\;
\begin{array}{l|ccc}
\mbox{in fm} & \;1/\omega_\psi\; &\; 1/\omega_{0^+}\; &\; 1/m_{0^+} \\\hline
\mbox{\sc r}=1     &  1.03  \; &\; 0.29\; &\; 0.31 \\
\mbox{\sc r}=0.65  &  0.95  \; &\; 0.28\; &\; 0.32
\end{array}\,.
\end{equation}
The scalar diquark parameter values determined in these unconstrained fits
are within $10$\% of those obtained in the BSE studies of Ref.\
[\ref{conradsepR}].

\begin{figure}[t]
\centering{\
\epsfig{figure=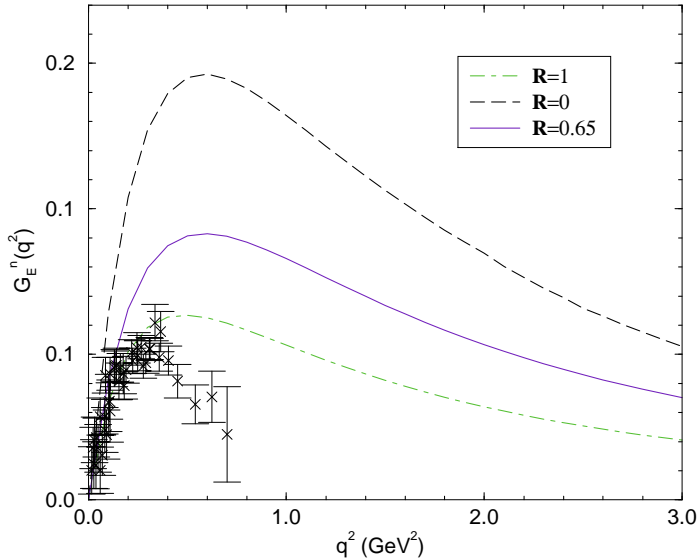,height=7.5cm}}
\parbox{33em}{\caption{\label{nnff} Neutron electric form factor calculated
with three different values of $\mbox{\sc r}$.  Data from
Ref.~[\protect\ref{saclayR}], extracted using the Argonne V18
potential.\protect\cite{bobpot} $G_E^n$ is very sensitive to the strength of
$f_2$.  (NB.\ The contribution of the pseudovector diquark was not included
in these calculations.)\hspace*{\fill}}}\vspace*{-2ex}
\end{figure}

Using the values in Eq.\ (\ref{paramsNFF}) a wide range of observables can be
calculated and herein we exemplify the results via
\begin{equation}
\label{results}
\begin{array}{l|ccccccc}
          & \;r_p^2\,({\rm fm}^2)\; & \;r_n^2\,({\rm fm}^2)\;&
          \;\mu_p (\mu_N)\; &\; \mu_n (\mu_N)\; &\;\mu_n/\mu_p & 
                g_{\pi NN} & g_A \\\hline
{\rm Emp.} & (0.87)^2 & -(0.34)^2 & 2.79 & -1.91 & -0.68 & 13.4 & 1.26
          \\\hline 
{\rm Calc.}_{\mbox{\sc r}=1}
           & (0.78)^2 & -(0.34)^2 & 2.82 & -1.62 & -0.57 & 14.6 & 1.27 \\ 
{\rm Calc.}_{\mbox{\sc r}=0.65}
           & (0.81)^2 & -(0.37)^2 & 2.85 & -1.63 & -0.57 & 14.5 & 1.12 \\ \hline
\end{array}
\end{equation}
and the neutron electric form factor in Fig.~\ref{nnff}.  One significant
feature, apparent in Eq.\ (\ref{results}), is that $|\mu_n/\mu_p| > 0.5$,
which is only possible because the impulse approximation explicitly includes
diquark breakup contributions.  Another is a prediction for the ratio $\mu_p
G_E^p(q^2)/G_M^p(q^2)$ that is in semi-quantitative agreement with recent
results from TJNAF.\cite{HallA} As shown explicitly in Ref.\
[\ref{revbastiR}], the impulse approximation requires five terms when the
diquark is composite, a fact also appreciated in Ref.\ [\ref{pichowskyR}].
It also requires an explicit form for the dressed-quark propagator but that
is well-known from studies of meson observables and is given; e.g., in Refs.\
[\ref{jacques1R},\ref{jacques2R}].

\begin{table}[t]
\begin{center}
\caption{\label{tableresults} Nucleon mass and $\mbox{\sc r}=f_2/f_1$ ratio
determined by solving the Fadde'ev equation.  $\omega_{\psi_{f_1}}$,
$\omega_{\psi_{f_2}}$ are the widths of a least-squares fit to $f_1$, $f_2$
assuming they are pointwise well-approximated by ${\cal
F}(\ell^2/\omega_{\psi_{f}})$.  [They actually fall faster with increasing
$k^2$ but these widths are nevertheless a useful guide for comparison with
Eq.\ (\protect\ref{results}).]  Rows $1$ and $5$ give the results obtained
with only a scalar diquark while the others were obtained with the inclusion
of a pseudovector diquark correlation.  All dimensioned quantities are given
in GeV.  In these calculations the dressed-quark's Euclidean
constituent-quark mass is $M_u^E=0.33\,$GeV, which
was fixed in independent studies of meson
observables.\protect\cite{conradkaon} Using the parameters in the last row,
an analogous Fadde'ev equation for the $\Delta(1232)$ yields a mass
$M_\Delta=1.23\,$GeV.
\hspace*{\fill}}\vspace*{1ex}
\begin{tabular}{c|cccc|cccc}
      & $\omega_{0^+}$ & $m_{0^+}$ & $\omega_{1^+}$ & $m_{1^+}$ & 
        {\sc r} & $\omega_{\psi_{f_1}}$ & $\omega_{\psi_{f_2}}$ & $M$
        \\\hline 
$0^+$ & $0.68$ & $0.64$ & -- & -- & $1.11$ & $0.40$ & $0.43$ & $1.48$   \\
$0^+$ \& $1^+$ & $0.68$ & $0.64$ & $0.68$ & $0.82$ 
                & $0.62$ & $0.38$ & $0.41$ & $1.28$ \\
$0^+$ \& $1^+$ & $0.68$ & $0.64$ & $0.40$ & $0.82$ 
                & $0.73$ & $0.32$ & $0.35$ & $1.16$ \\
 $0^+$ \& $1^+$ & $0.40$ & $0.64$ & $0.40$ & $0.82$ 
                 & $1.11$ & $0.29$ & $0.31$ & $1.14$ \\
$0^+$ \& $1^+$ & $0.68$ & $0.54$ & $0.40$ & $0.69$ 
                & $0.54$ & $0.31$ & $0.36$ & $0.94$ \\\hline
$0^+$ & $0.42$ & $0.64$ & -- & -- & $2.11$ & $0.34$ & $0.34$ & $1.31$   \\
$0^+$ \& $1^+$ & $0.42$ & $0.64$ & $0.42$ & $0.86$ 
                & $1.26$ & $0.31$ & $0.32$& $1.20$   \\
$0^+$ \& $1^+$ & $0.42$ & $0.64$ & $0.97$ & $0.86$ 
                & $1.68$ & $0.34$ & $0.33$& $1.27$   \\
$0^+$ \& $1^+$ & $0.42$ & $0.64$ & $1.09$ & $0.86$ 
                & $0.46$ & $0.41$ & $0.35$ & $0.94$ \\\hline
\end{tabular}
\end{center}
\vspace*{-2ex}
\end{table}

Hitherto the choice of $\mbox{\sc r}$ is arbitrary.  However, its value is
fixed by solving the Fadde'ev equation.  To exemplify that we have solved the
Fadde'ev equation, Eq.\ (\ref{faddeev}), with the values of $\omega_{0^+}$,
$m_{0^+}$ in Eq.\ (\ref{paramsNFF}) and assuming that
$f_i(\ell;P)=f_i(\ell^2)$.  This yields the results in Table
\ref{tableresults}.  As anticipated in Ref.\ [\ref{jacques2R}], a reduction
of $\sim 30\,$--$\,40\,$\% is required in the calculated mass for agreement with
experiment and here the inclusion of a pseudovector diquark can
help.\cite{reinhard}

To explore that we added such a correlation to Eq.\ (\ref{Psi3}):
\begin{equation}
\psi_3^{1^+}= \epsilon_{c_1 c_2 c_3}\, \Delta^{1^+}_{\mu\nu}(K)\,
i\gamma_\mu C \,\mbox{\tt t}^i\,\Gamma_{1^+}(\case{1}{2}p_{[12]})\,
{\cal A}^i_\nu(\ell;P)\,u(P)\,,
\end{equation}
where: $\{\mbox{\tt t}_{i=(+,0,-)}=
(\tau_0+\tau_3)/\sqrt{2},\tau_1,(\tau_0-\tau_3)/\sqrt{2}\}$, $\tau_0={\rm
diag}(1,1)$, are the symmetric isospin-triplet matrices,
\begin{equation}
{\cal A}^i_\nu(\ell;P)= a_1^i(\ell^2)\,\gamma_5\gamma_\nu + a_2^i(\ell^2)\,
\gamma_5\gamma\cdot\hat\ell\,\hat\ell_\nu \,,\;\;\hat \ell^2 = 1\,,
\end{equation}
with $a_1^1=a_1^2=a_1^3$, $a_2^1=a_2^2=a_2^3$ in the isospin symmetric limit;
and
\begin{eqnarray}
\Delta^{1^+}_{\mu\nu}(K) & = & \left(\delta_{\mu\nu}+\frac{K_\mu
K_\nu}{m_{1^+}^2}\right) \,\frac{1}{m_{1^+}^2}\,{\cal
F}(K^2/\omega_{1^+}^2)\,,\\
\Gamma_{1^+}(k^2) & = & \frac{1}{{\cal N}_{1^+}} {\cal
F}(k^2/\omega_{1^+}^2)\,;
\end{eqnarray}
and extended Eq.\ (\ref{faddeev}) to include the coupling to this channel.

The results obtained in this case are also presented in Table
\ref{tableresults}.  In the upper panel we chose the value of
$m_{1^+}/m_{0^+}$ from Ref.\ [\ref{conradsepR}] and, initially,
$\omega_{1^+}=\omega_{0^+}$, to find that the pseudovector diquark provides
additional attraction and reduces the calculated mass by $11$\%.  Reducing
$\omega_{1^+}$ increases the pseudovector diquark coupling via an increase in
$g_{1^+}:= 1/{\cal N}_{1^+}$, hence the calculated mass is reduced: a $40$\%
reduction in $\omega_{1^+}$ reduces $M$ by $10$\%.  Reducing $\omega_{0^+}$
by the same amount has very little effect, reducing $M$ by only an extra
$2$\%.  However, in agreement with intuition, decreasing the diquark masses
reduces the calculated nucleon mass: a $21$\% decrease yields a $19$\%
reduction in $M$.  In the lower panel, we expanded the system of coupled
equations to include an analogous Fadde'ev equation for the $\Delta$.  The
exploration of this system made clear that an increase in $g_{1^+}$ via a
decrease in $\omega_{1^+}$ is not the only way to reduce the nucleon mass.
Increasing $\omega_{1^+}$ also increases the support of the integrand that
describes the pseudovector's binding contribution and, for $\omega_{1^+}$
greater-than a certain value, that increase more than compensates for the
concomitant decrease in $g_{1^+}$, yielding increased binding and a lower
nucleon mass.  The table makes amply clear that an internally consistent
description of the nucleon and $\Delta$ is possible using only scalar and
pseudovector diquark correlations, just as found in Ref.\ [\ref{reinhardR}].
Furthermore, it is clear from the table that a well-constrained scalar
diquark model should employ \mbox{\sc r} in the range $\sim 0.5-0.7$.

A question that remains unaddressed is the role of pion loops.  The on-shell
$\pi NN$ coupling is strong and hence it is conceivable that such loops might
generate large self energy corrections to the nucleon's mass.  We have made
an estimate using a model DSE for the nucleon self energy:$\,$\footnote{The
positive sign before the integral on the r.h.s.\ is correct and entails that
both the vector and scalar self energies are positive.}
\begin{equation}
G^{-1}(p)  = i\gamma\cdot p + M + 
3\int\frac{d^4 k}{(2\pi)^4}\,\Delta_\pi(p-k)\,g^2_{\pi
NN}((p-k)^2,k^2) \,\gamma_5 G(k)\gamma_5\,,
\end{equation}
where $\Delta_\pi(\ell)= 1/(\ell^2+m_\pi^2)$ and $g_{\pi NN}((p-k)^2,k^2)$ is
the momentum-dependent $\pi NN$ coupling.  The $t=-(p-k)^2$-dependence of
this coupling was calculated for on-shell nucleons in Ref.\
[\ref{jacques2R}], with the result
\begin{equation}
\label{gpiNN}
g_{\pi NN}(t,k^2=-M^2)\approx \frac{g_{\pi NN}}{(1 - t/\Lambda_\pi^2)^2}\,,\;
\Lambda_\pi = 0.96\,{\rm GeV},
\end{equation}
where $g_{\pi NN}$ is given in Eq.\ (\ref{results}).  This is not quite
sufficient for our present purpose because the nucleon in the loop is not
on-shell.  Therefore to complete an estimate we employ a simple product {\it
Ansatz}: 
\begin{equation}
g_{\pi NN}(p;k)= \frac{g_{\pi NN}}{
(1 + |p^2+k^2|/\Lambda_\pi^2)^2\,
(1 + (|p^2+ M^2|+|k^2+ M^2|)/\Lambda_\pi^2)^2}\,,
\end{equation}
to approximate the {\it angular-average} of the coupling that is active in
the integral equation.  With the parameters in the last row of the upper
panel in Table \ref{tableresults} we find that the pion loop adds $10\,$MeV
to the nucleon's mass; i.e., it provides only a $1$\% increase.  (The
detailed form of Eq.\ (\ref{gpiNN}) is not important but the off-shell
suppression is.)  In addition, if we define $M_{m_\pi}$ to be the
loop-corrected nucleon mass obtained using a particular value of $m_\pi$,
then $M_{m_\pi}$ decreases with increasing $m_\pi$, in qualitative agreement
with chiral perturbation theory.  (See; e.g., Ref.\ [\ref{ugmR}].)

\medskip

\hspace*{-\parindent}{\bf Epilogue.}~~Two short summaries are all we have
room for here.  In large part, the light-quark meson sector is well
understood.  The renormalisation-group-improved rainbow-ladder truncation
provides reliable information in many channels and where it doesn't the
reasons why are understood.\cite{truncscheme} Understanding the nucleon: it's
mass and interactions, is a contemporary focus and progress is rapid.
Success there will open the way for many new phenomenological applications,
such as the calculation of quark distribution functions, which are merely
parametrised in analyses of deep inelastic scattering.  That the DSEs can
provide valuable insight here is demonstrated by a calculation of the valence
quark distribution in the pion,\cite{uvx} which itself is measurable given a
high-luminosity electron-proton collider.\cite{roy} Analogous, detailed
Fadde'ev-equation-based studies of the $\Delta$ are also
beginning, as illustrated herein and in Ref.\ [\ref{reinhardDeltaR}].

\medskip

\hspace*{-\parindent}{\bf Acknowledgments.}~~CDR is grateful for the
hospitality and support of the Erwin Schr\"odinger Institute for Mathematical
Physics, Vienna, which helped to make possible his participation in this
workshop, and to the organisers of Quark Confinement and the Hadron Spectrum
IV; SMS is grateful for financial support from the A.\ v.\ Humboldt
foundation; and we acknowledge useful communications with R.~Alkofer,
J.C.R.~Bloch, P.~Maris, D.R.~Phillips and P.C.~Tandy.  This work was
supported by the US Department of Energy, Nuclear Physics Division, under
contract no. W-31-109-ENG-38, the National Science Foundation under grant
no.\ INT-9603385, and benefited from the resources of the National Energy
Research Scientific Computing Center.

\centerline{\rule{30em}{0.1ex}}
\begin{flushleft}
 
%
\end{flushleft} 
\end{document}